\documentclass[12pt,reqno]{amsart}
\usepackage{amsmath,amsfonts,amsthm,amssymb}
\newcommand\version{October 24, 2006}
\usepackage{amsxtra}

\font\notefont=cmsl8  
\newtheorem{theorem}{Theorem}[section]
\newtheorem{proposition}[theorem]{Proposition}
\newtheorem{lemma}[theorem]{Lemma}
\newtheorem{corollary}[theorem]{Corollary}

\theoremstyle{definition}

\theoremstyle{remark}
\newtheorem{remark}[theorem]{Remark}

\numberwithin{equation}{section}

\newcommand{\C}{\mathbb{C}}

\newcommand{\eps}{\epsilon}
\renewcommand{\epsilon}{\varepsilon}

\renewcommand{\phi}{\varphi}
\newcommand{\R}{\mathbb{R}}

\DeclareMathOperator{\supp}{supp}

\DeclareMathOperator{\Tr}{Tr}
\DeclareMathOperator{\tr}{Tr}

\begin{document}

\title[Stability of Relativistic Matter --- \version]{Stability of
     Relativistic Matter with \\ Magnetic Fields for Nuclear Charges \\up
     to the Critical Value}

\author{Rupert L. Frank}
\address{Rupert L. Frank, Department of Mathematics, Royal Institute
    of Technology, 100 44 Stockholm, Sweden}
\email{rupert@math.kth.se}

\author{Elliott H. Lieb}
\address{Elliott H. Lieb, Departments of Mathematics and Physics,
Princeton University,
    P.~O.~Box 708, Princeton, NJ 08544, USA}
\email{lieb@princeton.edu}

\author{Robert Seiringer}
\address{Robert Seiringer, Department of Physics, Princeton University,
P.~O.~Box 708,
    Princeton, NJ 08544, USA}
\email{rseiring@princeton.edu}

\thanks{\copyright\, 2006 by the authors. This paper may be reproduced, in its
entirety, for non-commercial purposes.}

\begin{abstract}
     We give a proof of stability of relativistic
     matter with magnetic fields all the way up to the critical value of
     the nuclear charge $Z\alpha=2/\pi$.
\end{abstract}

\date{\version}
\maketitle



\section{Introduction}

We shall give a proof of the `stability of relativistic matter'
that goes further than previous proofs by permitting the
inclusion of magnetic fields for values of the nuclear charge $Z$ all the
way up to $Z\alpha = 2/\pi$, which is the well known critical value in
the absence of a field.  (The dimensionless number $\alpha = e^2/\hbar
c$ is the `fine-structure constant' and equals 1/137.036  in nature.) More
precisely, we shall show how to modify the earlier proof of Theorem~2
in \cite{LY} so that an arbitrary magnetic field can be
included. Reference will freely be made to items in the \cite{LY}
paper.

The quantum mechanical Hamiltonian used here and in \cite{LY}, as well
as the definition of stability of matter, will be given in the next
section. For a detailed overview of this topic, we refer to \cite{L1,
  L2}.  For the present we note that stability requires a bound on
$\alpha$ in two ways. One is the requirement, for any number of
electrons, that $Z\alpha \leq 2/\pi$. In fact, if $Z\alpha >2/\pi$ the
Hamiltonian is not bounded below even for a single electron. The other
requirement is a bound on $\alpha$ itself, $\alpha \leq \alpha_c$,
even for arbitrarily small $Z>0$, which comes into play when the
number of particles is sufficiently large. It is known that
$\alpha_c\leq 128/15\pi$; see \cite[Thm.~3]{LY} and also \cite{DL}.

For values of $Z\alpha$ strictly smaller than the critical value
$2/\pi$, it has been shown that stability holds with a magnetic field
included. This is the content of Theorem~1 in \cite{LY}, in which the
critical value of $\alpha_c$ goes to zero as $Z\alpha$ approaches
$2/\pi$, however. (The result in \cite[Theorem 1]{LY} does not
explicitly include a magnetic field, but the fact that the proof can
easily be modified was noted in \cite{LLSo}.)  A similar result, by a
different method, was proved in \cite{LLSi}.

The more refined Theorem 2 in \cite{LY} gives stability for the
`natural' value $Z\alpha \leq 2/\pi$  and all $\alpha \leq 1/94$. While the true
value of $\alpha_c$ is probably closer to 1, the value $1/94 > 1/137$
is sufficient for physics.  The problem with the proof of \cite[Theorem
2] {LY} is that it does not allow for the inclusion of  magnetic fields.  Specifically,
Theorems 9--11 have to be substantially modified, and doing so was an
open problem for many years.  This will be accomplished here at the
price of decreasing  $\alpha_c$ from 1/94 to 1/133. Fortunately,
this is still larger than the physical value 1/137 !

In a closely related paper \cite{FLS} we also show how to achieve a
proof of stability for all $Z\alpha\leq 2/\pi$ with an arbitrary
magnetic field, but the value of $\alpha_c$ there is very much smaller
than the value obtained here. In particular, the physical value of
$\alpha= 1/137$ is not covered by the result in \cite{FLS}. The focus
of \cite{FLS} is much broader than `stability of matter', however. It
is concerned with a general connection between Sobolev and
Lieb-Thirring type inequalities, and includes as a special case
Theorem~\ref{thm:lth} of this paper. The proof of the general result
in \cite{FLS} is much
more involved than the one of the special case presented here, and
yields a worse bound on the relevant constant.

\bigskip {\it Acknowledgements.} We thank Heinz Siedentop for helpful
remarks. This work was partially supported by the Swedish Foundation
for International Cooperation in Research and Higher Education (STINT)
(R.F.), by U.S.  National Science Foundation grants PHY 01 39984
(E.L.) and PHY 03 53181 (R.S.), and by an A.P. Sloan Fellowship
(R.S.).


\section{Definitions and Main Theorem}

We consider $N$ electrons of mass $m\geq 0$ with $q$ spin states
($q=2$ for real electrons) and $K$ fixed nuclei with (distinct)
coordinates $R_1,\ldots,R_K\in\R^3$ and charges $Z_1,\ldots,Z_K> 0$.
The electrons interact with an external, spatially dependent magnetic
field $B(x)$, which is given in terms of the magnetic vector potential
$A(x)$ by $B={\rm curl} A$.  A pseudo-relativistic description of the
corresponding quantum-mechanical system is given by the Hamiltonian
\begin{equation}\label{def:ham}
     H_{N,K} := \sum_{j=1}^N \left(\sqrt{(p_j+ A(x_j))^2+m^2}-m\right)
     + \alpha V_{N,K}(x_1,\dots,x_N;R_1,\dots,R_k)\,.
\end{equation}
The Pauli exclusion principle for fermions dictates that $H_{N,K}$
acts on functions in the anti-symmetric $N$-fold tensor
product $\wedge^N L^2(\R^3;\C^q)$. We use units in which
$\hbar=c=1$, $\alpha>0$ is the fine structure constant, and
\begin{align}
     V_{N,K}(x_1,\dots,x_N; R_1, \dots
R_K) := & \sum_{1\leq i<j\leq N} |x_i-x_j|^{-1}
     - \sum_{j=1}^N \sum_{k=1}^K Z_k |x_j-R_k|^{-1} \notag \\
     & + \sum_{1\leq k<l\leq K} Z_k Z_l |R_k-R_l|^{-1}.
\end{align}
is the Coulomb potential (electron-electron, electron-nuclei,
nuclei-nuclei, respectively).  In this model there is no interaction
of the electron {\it spin} with the magnetic field. Note that we
absorb the electron charge $\sqrt{\alpha}$ into the vector potential
$A$, i.e., we write $A(x)$ instead of $\sqrt{\alpha}A(x)$ in
(\ref{def:ham}).  Since $A$ is arbitrary and our bounds are
independent of $A$, this does not affect our results.

{\it Stability of matter} means that $H_{N,K}$ is bounded from below by a
constant times $(N+K)$, independently of the positions $R_k$ of the nuclei
and of $A$. For a thorough discussion see \cite{L1, L2}.
By scaling all spatial coordinates it is easy to see that either
$\inf_{R_k, A}({\rm inf\, spec\,} H_{N,K}) = -mN$ or $=-\infty$.

We shall prove the following.

\begin{theorem}[{\bf Stability of relativistic matter with magnetic fields}]
\label{stability}
     For  $q\alpha\leq 1/66.5$ and $\alpha Z_j \leq 2/\pi$ for all $j$,
     \begin{equation*}
         H_{N,K} \geq -mN
     \end{equation*}
for all $N$, $K$, $R_1,\dots,R_K$ and $A$.
\end{theorem}

For electrons $q=2$ and hence our proof works up to
$$
\alpha = \frac 1{133} > \frac 1{137}\,.
$$

The rest of this paper contains the proof of Theorem \ref{stability}, but
let us
first state an obvious fact.

\begin{corollary}
As a multiplication  operator on
$\wedge^N L^2 (\R^3;\C^q) $,
\begin{equation} \label{ineq}
V_{N,K}(x_1,\dots,x_N; R_1, \dots R_K) \geq - \max\{66.5\,q,\  \pi Z_j/2 \}\,
\sum_{j=1}^N |p_j+ A(x_j)|
\end{equation}
for all $A$.
\end{corollary}

This, of course, is just a rewording of Theorem \ref{stability}, but
the point is that it provides a lower bound for the Coulomb potential
of interacting particles in terms of a one-body operator $|p +A(x)|$.
This operator is dominated by the \emph{nonrelativistic} operator
$|p+A(x)|^2$ and, therefore, \eqref{ineq} is useful in certain
nonrelativistic problems.  For example, an inequality of this type was
used in \cite{LLSo} to prove stability of matter with the Pauli
operator $|p+A(x)|^2 +\sigma \cdot B(x)$ in place of $|p+A(x)|^2$. It
was also used in \cite{LSiSo} to control the no-pair Brown-Ravenhall
relativistic model.

An examination of the proof of Theorem~2 in \cite{LY} shows that there
are two places that do not permit the inclusion of a magnetic vector
potential $A$. These are Theorem~9 (Localization of kinetic energy --
general form) and Theorem~11 (Lower bound to the short-range energy in
a ball). Our Theorem~\ref{imsmag} is precisely the extension of
Theorem~9 to the magnetic case. It may be regarded as a diamagnetic
inequality on the localization error.  It implies that Theorem~10 in
\cite{LY} holds also in the magnetic case, without change except for
replacing $|p|$ by $|p+A|$; see Theorem \ref{kinloc} below.

A substitute for Theorem~11 in \cite{LY} will be given in Theorem
\ref{thm:lth} below. It is based on the observation that an estimate
on eigenvalue sums of a non-magnetic operator with discrete spectrum
implies a similar estimate (with a modified constant) for the
corresponding magnetic operator.  This is not completely obvious,
since there is no diamagnetic inequality for \emph{sums} of
eigenvalues. (In fact, a conjectured diamagnetic inequality actually
fails for fermions on a lattice and leads to the `flux phase'
\cite{L3}.) It is for the different constants in Theorem~11 in
\cite{LY} and in our Theorem~\ref{thm:lth} that our bound on
$\alpha_c$ become worse than the one in \cite{LY}.

As should be clear from the above discussion, our main tool will be a
diamagnetic inequality for single functions. The one we use is the
diamagnetic inequality for the heat kernel. In the relativistic case it
states that for any $A\in L^2_{\rm loc}(\R^3;\R^3)$ and $u\in L^2(\R^3)$
one has
\begin{equation}\label{eq:diamag}
\left| \big(\exp(-t|p+A|) u \big)(x) \right|
\leq \big( \exp(-t|p|) |u| \big) (x),
\quad x\in\R^3.
\end{equation}
This follows with the help of the subordination formula
$$
e^{-|\xi|} = \int_0^\infty e^{-t-|\xi|^2/(4t)}  \,
\frac{dt}{\sqrt{\pi t}}\,
$$
from the `usual' (nonrelativistic) diamagnetic inequality for the
semigroup $\exp(-t|p+A|^2)$; see, e.g., \cite{Si2}. The heat kernel is
not prominent in \cite{LY}, and our reformulation of some of the key
estimates in \cite{LY} in terms of the heat kernel is the
principal novel feature of this paper.


\section{Localization of the kinetic energy with magnetic fields}

\subsection{Relativistic IMS formula}

In this subsection we establish the analogue of Theorem~9 in \cite{LY}
in the general case $A\not=0$. First, recall that the IMS formula in the
nonrelativistic case says that for any $u$ and $A$
\begin{equation*}
     \int_{\R^3} |(p+A)u|^2 \,dx
     = \sum_{j=0}^n \int_{\R^3}
     |(p+A) (\chi_j u)|^2 \,dx
     - \int_{\R^3} \sum_{j=0}^n |\nabla \chi_j|^2 |u|^2 \,dx
\end{equation*}
whenever $\chi_j$ are real functions with $\sum_{j=1}^n\chi_j^2\equiv 1$.
In this case the localization error $\sum_{j=0}^n |\nabla \chi_j|^2$ is
local and independent of $A$. The analogue in the relativistic case is
the following special case of \cite[Lemma~B.1]{FLS}. For
the sake of completeness, we include its proof here.

\begin{theorem}[{\bf Localization of kinetic energy -- general
form}]\label{imsmag}
     Let $A\in L^2_{\rm loc}(\R^3;\R^3)$. If $\chi_0,\ldots,\chi_n$ are
real Lipschitz continuous functions
     on $\R^3$ satisfying $\sum_{j=0}^n \chi_j^2 \equiv 1$, then one has
     \begin{equation}\label{eq:imsmag}
       (u, |p+A| u) = \sum_{j=0}^n (\chi_j u, |p+A| \chi_j u) - (u,L_A u)\,.
     \end{equation}
     Here $L_A$ is a bounded operator with integral kernel
     \begin{equation*}
       L_A(x,y) := k_A(x,y) \sum_{j=0}^n (\chi_j(x)-\chi_j(y))^2\,,
     \end{equation*}
     where $k_A(x,y) := \lim_{t\uparrow 0} t^{-1}\exp(-t|p+A|)(x,y)$ for
a.e. $x,y\in\R^3$ and
     \begin{equation}\label{eq:kak}
       |k_A(x,y)| \leq \frac{1}{2\pi^2 |x-y|^{4}}\,.
     \end{equation}
\end{theorem}

Note that (\ref{eq:kak}) says that
\begin{equation}\label{eq:defL}
|L_A(x,y)| \leq L(x,y) : = \frac{1}{2\pi^2|x-y|^{4}}  \sum_{j=0}^n
(\chi_j(x)-\chi_j(y))^2\,.
\end{equation}
Here, $L(x,y)$ is the same as in \cite[Eq.~(3.7)]{LY}. Therefore,
\eqref{eq:kak} is
a diamagnetic inequality for the localization error.

\begin{proof}
     We write $k_A(x,y,t) := \exp(-t|p+A|)(x,y)$ for the heat kernel
     and find
     \begin{align*}
       & \sum_{j=0}^n  (\chi_j u, (1-\exp(-t|p+A|)) \chi_j u)
       = (u, (1-\exp(-t|p+A|)) u)\\
       & \qquad\qquad  + \frac 12\sum_{j=0}^n
       \iint k_A(x,y,t)(\chi_j(x)-\chi_j(y))^2 \overline{u(x)} u(y)
       \,dx\,dy\,.
     \end{align*}
     (This is proved simply by writing out both sides in terms of
$k_A(x,y,t)$ and using $\sum\chi_j^2 \equiv 1$.)
     Now we divide by $t$ and let $t\to 0$. The left side
     converges to $\sum_{j=0}^n (\chi_j u, |p+A| \chi_j u)$.  Similarly,
     the first term on the right side divided by $t$ converges to
     $(u,|p+A| u)$. Hence the last term divided by $t$ converges to
     some limit $(u, L_A u)$. The diamagnetic inequality \eqref{eq:diamag}
     says that
     \begin{equation*}
     |k_A(x,y,t)| \leq \exp(-t |p|)(x,y)
     = \frac t{\pi^2\left(|x-y|^2+t^2\right)^2} 
     \end{equation*}
     (see \cite[Eq.~7.11(9)]{LL}). This implies, in particular,
     that $L_A$ is a bounded operator. Now it is easy to check that
     $L_A$ is an integral operator and that the absolute value of its
     kernel is bounded pointwise by the one of $L$ in (\ref{eq:defL}).
\end{proof}


\subsection{Localization of the kinetic energy}

In this subsection we will bound the
localization error $L_A$ by a potential energy correction and an
additive constant. This is the extension of Theorem~10 in
\cite{LY} to the case $A\neq0$. It is important that both error
terms in our bound can be chosen independently of $A$.

First we need to introduce some notation. We write
\begin{equation*}
        \mathcal B_R := \{x\, :\, |x|<R\}
\end{equation*}
for the ball of radius $R$ and $\chi_{\mathcal B_R}$ for its
characteristic function. If $R=1$, we omit the index in the notation. We
fix a constant $0<\sigma<1$
and Lipschitz continuous functions $\chi_0,\chi_1$ with
$\chi_0^2+\chi_1^2\equiv 1$ such that
$\supp\chi_1\subset\overline{\mathcal B_{1-\sigma}}$.
With these we define $L$ as in (\ref{eq:defL}) with $n=1$.  We decompose
$L$ in a short-range part $L^0$ and a long-range part $L^1$ given by
the kernels
\begin{equation}\label{eq:decompl}
     L^1(x,y) := L(x,y)
     \chi_{\mathcal B}(x) \chi_{\mathcal B}(y) \chi_{\mathcal
     B_\sigma}(x-y),
     \qquad
     L^0(x,y) := L(x,y) -  L^1(x,y).
\end{equation}
Define
\begin{equation}\label{eq:omega}
     \Omega := \frac12 \tr\left(L^0\right)^2
\end{equation}
and, for an arbitrary positive function $h$ on $\mathcal B$,
\begin{equation*}\label{eq:theta}
     \theta(x) := h^{-1}(x) \int_\mathcal B L^1(x,y) h(y)\,dy
     = h^{-1}(x) \chi_{\mathcal B}(x)
     \int_{|y|<1,\,|x-y|<\sigma} L(x,y) h(y)\,dy \,.
\end{equation*}
Finally, for $\epsilon>0$ we define the function
\begin{equation}\label{eq:uepsilon}
     U_{\epsilon}^* :=
     \epsilon \chi_{\mathcal B_{1-\sigma}} + \theta
\end{equation}
and note that $U_{\epsilon}^*$ is supported in $\overline{\mathcal B}$.

\begin{theorem}[{\bf Localization of kinetic energy -- explicit bound in the
    one-center case}]\label{kinloc} For any $\epsilon>0$ and any
  non-negative trace-class operator $\gamma$ one has
     \begin{equation}\label{eq:kinloc}
       \tr\gamma|p+A| \geq
       \sum_{j=0}^1 \tr\chi_j\gamma\chi_j(|p+A|-U_{\epsilon}^*)
       - \epsilon^{-1} \Omega\|\gamma\|.
     \end{equation}
\end{theorem}

For $A=0$ this is exactly Theorem 10 in \cite{LY}. As explained there,
$U_{\epsilon}^*$ is a potential energy correction with only slightly
larger support than $\chi_1$. The last term in \eqref{eq:kinloc} is
due to the long range nature of $|p+A|$. It depends on $\gamma$
through its norm $\|\gamma\|$ but not through its trace.
We emphasize again that both error terms in the inequality
\eqref{eq:kinloc} are independent of $A$.

\begin{proof}
     The localization formula \eqref{eq:imsmag} yields
     \begin{equation*}
       \tr\gamma|p+A| =
       \sum_{j=0}^1 \tr\chi_j\gamma\chi_j|p+A|
       - \tr \gamma L_A,
     \end{equation*}
     so we only have to find an upper bound for $\tr\gamma L_A$. We
     decompose $L_A= L^0_A + L^1_{A}$ in the manner of \eqref{eq:decompl}
and,
     following the proof of Theorem 10 in \cite{LY} word by word, we
     obtain
     \begin{equation*}
       \tr\gamma L^0_A \leq
       \epsilon \tr\gamma\chi_{\mathcal B_{1-\sigma}}
       + (2\epsilon)^{-1} \|\gamma\| \tr\left(L^0_A\right)^2,
       \qquad
       \tr\gamma L^1_{A} \leq \tr\gamma\theta_{A}.
     \end{equation*}
     Here $\theta_{A}(x):=0$ if $x\not\in\mathcal B$ and, if
     $x\in\mathcal B$,
     \begin{equation*}
       \theta_{A}(x) :=
       h^{-1}(x) \int_\mathcal B |L_{A}^1(x,y)| h(y)\,dy.
     \end{equation*}
     The estimate $|L_A(x,y)| \leq L(x,y)$ from Theorem~\ref{imsmag}
     implies that $\tr\left(L^0_A\right)^2 \leq 2\Omega$ and that
     $\theta_{A} \leq \theta$. This leads to the stated lower bound.
\end{proof}


\section{Bounds on eigenvalues in balls}

So far we have considered $|p+A|$ and its heat kernel. Now we address
$|p+A|-2/(\pi|x|)$ and its heat kernel.
First of all, let us recall Kato's inequality \cite[Eq.~(V.5.33)]{K}
\begin{equation}\label{eq:kato}
(u,|p|u) \geq (2/\pi)(u,|x|^{-1} u)\,.
\end{equation}
(See also \cite{H,W,KPS}.)

Now let $\Gamma\subset \R^3$ be an open set (we shall be interested in the
case where $\Gamma$ is a ball) and consider the quadratic form given
by $Q_\Gamma (u)= (u, (|p|-2/\pi|x|)u)$, restricted to those functions
$u\in L^2 (\R^3) $ that satisfy $u=0$ on $\Gamma^c$, the complement of
$\Gamma$. Of course, we also require $u $ to be in the quadratic form
domain of $|p|-2/\pi|x|$. The quadratic form $Q_\Gamma$ is non-negative by
\eqref{eq:kato}
and it is closed (because the form $|p|-2/\pi|x|$
is closed on $L^2(\R^3) $ and limits of functions that are zero on
$\Gamma^c$ are zero on $\Gamma^c$). From this it follows that there is
a self-adjoint operator $H_\Gamma $ on some domain in $L^2(\Gamma)$
such that $Q_\Gamma (u) = (u, H_\Gamma u)$.
With this operator, we
can define the `heat kernel' $\exp\left(-tH_\Gamma\right)$ on $L^2(\Gamma)$
and its trace. (The fact that the trace is finite when the volume of
$\Gamma$ is finite follows from subsequent considerations.)

Similarly, for a magnetic vector potential $A\in L^2_{\rm loc}(\R^3;\R^3)$,
we define the operator $H^A_\Gamma$
in $L_2(\Gamma)$ using the quadratic form
$ (u, (|p+A|-2/\pi|x|)u)$. Note that \eqref{eq:diamag} implies that
\begin{equation}\label{eq:diamag1}
(u,|p+A| u)\geq (|u|,|p| |u|) \,.
\end{equation}
This, together with \eqref{eq:kato}, shows that $(u, (|p+A|-2/\pi|x|)u)$
is non-negative.

\begin{lemma}[{\bf Heat kernel diamagnetic inequality}]\label{lemma:dia}
Let $\Gamma\subset \R^3$ and let $A\in L^2_{\rm
loc}(\R^3;\R^3)$.
Then, for any $t>0$,
\begin{equation}\label{eq:dia}
{\rm Tr}_{L^2(\Gamma)} \exp\left( -t H^A_\Gamma\right)
\leq {\rm Tr}_{L^2(\Gamma)} \exp\left( -t H_\Gamma\right)\,.
\end{equation}
\end{lemma}

\begin{proof}
    For $n=0,1,2,\ldots$ let $h_n := |p| - 2/(\pi|x|) + n \chi_{\Gamma^c}$
in $L^2(\R^3)$,
    where $\chi_{\Gamma^c}$ denotes the characteristic function of the
    complement of $\Gamma$. Similarly, let $h_n^A := |p+A| -2/(\pi|x|) +
    n\chi_{\Gamma^c}$. The diamagnetic inequality \eqref{eq:diamag}
    and standard approximation arguments
using Trotter's product formula imply that, for any
    $u\in L^2(\R^3)$,
$$
\big|\big(\exp(-t h^A_n)u \big)(x)\big| \leq  \big(\exp(-t h_n)|u|\big)(x)\,.
$$
(See \cite[Section~6.2]{FLS} for details of the argument.)

By the monotone convergence theorem \cite[Thm.~4.1]{Si}, $\exp(-t h_n)$
converges
strongly to $\exp(-t H_\Gamma)$ on the subspace $L^2(\Gamma)$, and
similarly for $h_n^A$. It follows that, for any $u\in L^2(\Gamma)$,
$$
\big|\big(\exp(-t H^A_\Gamma)u\big)(x)\big| \leq  \big(\exp(-t
H_\Gamma)|u|\big)(x)\,.
$$
Theorem~2.13 in \cite{Si2} yields the inequality $\| \exp(-t
H^A_\Gamma)\|_2\leq \| \exp(-t H_\Gamma)\|_2$ for the Hilbert-Schmidt
norm, and hence $\| \exp(-2t H^A_\Gamma)\|_1\leq \| \exp(-2t
H_\Gamma)\|_1$ for the trace norm by the semigroup property. This holds
for all $t>0$, and hence proves (\ref{eq:dia}).
\end{proof}

We use the notation $(x)_- = \max\{0,-x\}$ for the negative part of
$x\in\R$ in the following.

\begin{lemma}\label{lemma}
Assume that there is constant $M>0$ such that
\begin{equation}\label{eq:ass}
{\rm Tr}_{L^2(\Gamma)} \left( H_\Gamma -
     \Lambda \right)_- \leq M \Lambda^4
\end{equation}
for all $\Lambda\geq 0$. Then
\begin{equation}\label{eq:lem}
{\rm Tr}_{L^2(\Gamma)} \left( H^A_\Gamma -
     \Lambda \right)_- \leq \frac{6 e^3}{4^3} M \Lambda^4
\end{equation}
for all $\Lambda\geq 0$.
\end{lemma}

We note the the numerical factor in (\ref{eq:lem}) equals $6(e/4)^3
\approx 1.883$. This factor is the price we have to pay, using our
methods, to include an arbitrary magnetic field. It is the reason of the
decrease of $\alpha_c$ from $1/94$ to $1/133$.

\begin{proof}
Since $(x)_- \leq e^{-x-1}$, we have
$$
{\rm Tr}_{L^2(\Gamma)} \left( H^A_\Gamma-\Lambda\right)_-
\leq \frac{e^{t\Lambda}}{te} {\rm Tr}_{L^2(\Gamma)} \exp\left( -t
H_\Gamma^A\right)
$$
for any $t>0$. Using the diamagnetic inequality (\ref{eq:dia}),
$$
{\rm Tr}_{L^2(\Gamma)} \exp\left( -t H^A_\Gamma\right)
\leq {\rm Tr}_{L^2(\Gamma)} \exp\left( -t H_\Gamma\right)\,.
$$
Moreover, integrating by parts twice, $e^{-tx} = t^2 \int_0^\infty
e^{-t\lambda} (x-\lambda)_-\, d\lambda$, and hence
$$
{\rm Tr}_{L^2(\Gamma)} \exp\left( -t H_\Gamma\right)
= t^2 \int_0^\infty e^{-t\lambda}\, {\rm
     Tr}_{L^2(\Gamma)} \left( H_\Gamma  - \lambda
\right)_- \, d\lambda\,.
$$
Using the assumption (\ref{eq:ass}), we thus obtain
$$
{\rm Tr}_{L^2(\Gamma)} \left( H_\Gamma^A-\Lambda\right)_- \leq \frac{t
e^{t\Lambda}}{e} M
\int_0^\infty e^{-t\lambda} \lambda^4  d\lambda =
24\frac{e^{t\Lambda}}{t^4 e} M\,.
$$
To minimize the right side, the optimal choice of $t$ is
$t=4/\Lambda$. This yields (\ref{eq:lem}).
\end{proof}

In \cite[Thm.~11]{LY} it is shown that (\ref{eq:ass}) holds for
$\Gamma=\mathcal B_R$ a ball of radius $R$ centered at the origin. More
precisely, the following proposition
holds.

\begin{proposition}\label{prop}
For any $R> 0$ and $\Lambda\geq 0$,
\begin{equation*}
{\rm Tr}_{L^2(\mathcal B_R)} \left( H_{\mathcal B_R} -
     \Lambda \right)_- \leq  4.4827 \, R^3 \Lambda^4 \,.
\end{equation*}
\end{proposition}

Proposition~\ref{prop} follows from Theorem~11 in \cite{LY} by
choosing $\chi$ to be the characteristic function of the ball
$\mathcal B_R$, $q=1$ and $\gamma$ to be the projection onto the negative
spectral
subspace of $H_{\mathcal B_R} -\Lambda$.

\begin{remark}
It is illustrative to compare Proposition~\ref{prop} with the
Berezin-Li-Yau type
bound
\begin{equation}\label{eq:bly}
{\rm Tr}_{L^2(\Gamma)} \left( |p|_\Gamma  -
     \Lambda \right)_- \leq  \frac{1}{(2\pi)^3} \int_{\R^3} \int_\Gamma
\left( |\xi| - \Lambda\right)_- \, dx\, d\xi = \frac 1{24\pi^2}
\Lambda^4 |\Gamma| \,.
\end{equation}
(This can be proved in the same way as \cite[Thm.~12.3]{LL}.) 
The right side of (\ref{eq:bly}) is the semi-classical phase-space integral. The
operator $|p|_\Gamma$ is defined as $H_\Gamma$ above, but without the
Hardy-term $2/(\pi|x|)$. If the Hardy term were added, the phase-space
integral
would diverge (provided $\Gamma$ contains the origin), but
Proposition~\ref{prop}
says that a bound of the  form (\ref{eq:bly}) still holds.
(An examination of the proof in \cite{LY} shows that
Proposition~\ref{prop} actually holds for any open set $\Gamma$ of finite
measure.)
\end{remark}

Combining Lemma~\ref{lemma} and Proposition~\ref{prop} we obtain the
following theorem, which replaces
\cite[Thm.~11]{LY} in the magnetic case.

\begin{theorem}[{\bf Lower bound on the short-range energy in a
ball}]\label{thm:lth}
Let $C>0$ and $R>0$ and let
$$
H^A_{CR} := |p+A| - \frac{2}{\pi |x|} - \frac CR
$$
be defined on $L^2(\R^3)$ as a quadratic form. Let $0\leq
\gamma\leq q$ be a density matrix (i.e., a positive trace-class
operator) and let $\chi$ by any bounded function with support in
$\mathcal B_R$. Then
\begin{equation}
\Tr \overline \chi \gamma\chi H^A_{CR} \geq - 8.4411 \frac{q C^4 }{R}
\|\chi\|_\infty^2 \,.
\end{equation}
\end{theorem}

As compared with \cite[Thm.~11]{LY}, the constant has been multiplied
by $6(e/4)^3$, and $\|\chi\|_\infty^2$ appears instead of $|\mathcal
B_R|^{-1} \|\chi\|_2^2$.

\begin{proof}
Note that
$$
\Tr \overline \chi \gamma\chi H^A_{CR}
= \Tr  \overline \chi \gamma\chi\left( H_{\mathcal B_R}^A - C/R\right)
\geq - \| \overline\chi\gamma\chi\|_\infty \, \Tr_{L^2(\mathcal B_R)} (
H_{\mathcal B_R}^A -C/R)_-\,.
$$
The assertion follows from Lemma~\ref{lemma} and
Proposition~\ref{prop}, observing that $
\|\overline\chi\gamma\chi\|_\infty \leq q
\|\chi\|_\infty^2$.
\end{proof}


\section{Proof of Theorem \ref{stability}}

We assume that the reader is familiar with the proof of Theorem 2 in
\cite{LY}. We shall only emphasize changes in their argument. The
main idea is to replace Theorems 10 and 11 in \cite{LY} by our
Theorems~\ref{kinloc} and~\ref{thm:lth}, respectively.

There are some immediate simplifications. First, in view of the simple
inequality $\sqrt{|p|^2+m^2}\geq |p|$ it is enough to prove the
Theorem~\ref{stability} for $m=0$. Moreover, by the convexity argument of \cite{DL} it
suffices to treat the case $Z_1=\ldots=Z_K=:z$ and $\alpha z = 2/\pi$.
So henceforth we assume $m=0$, $Z_1=\ldots=Z_K=z$ and $\alpha z=2/\pi$.

Let $D_k := \min\{|R_k-R_l| :
l\not =k\}$ and define the Voronoi cell
\begin{equation*}
     \Gamma_k :=
     \{x\in\R^3 : |x-R_k|<|x-R_l| \textrm{ for all $l\not =k$} \}.
\end{equation*}
Fix $0<\lambda<1$ and define a function $W:=G+F$ in each Voronoi cell
by
\begin{equation*}
     G(x) := z|x-R_k|^{-1},
     \qquad
     F(x) := D_k^{-1} \tilde F(|x-R_k|/D_k),
     \qquad x\in\Gamma_k,
\end{equation*}
where
\begin{equation*}
     \tilde F(t) := \left\{
     \begin{array}{ll}
       2^{-1} (1-t^2)^{-1}
       & \textrm{if $t\leq\lambda$}, \\
       (\sqrt{2z} + \frac12)t^{-1}
       & \textrm{if $t>\lambda$}.
     \end{array} \right.
\end{equation*}
By the electrostatic inequality in \cite[Sect.~III, Step A]{LY} our
Theorem \ref{stability} will follow if we can prove that
\begin{equation}\label{eq:stabilityoneparticle}
     \tr\gamma(|p+A| - \alpha W)
     \geq -\frac{z^2\alpha}8 \sum_{k=1}^K D_k^{-1}
\end{equation}
for some $0<\lambda<1$ and all density matrices $\gamma$ with
$0\leq \gamma\leq q$). Note that \eqref{eq:stabilityoneparticle} is an
inequality for a \emph{one-particle} operator.

For fixed $0<\sigma<1/3$ we choose $\chi$, $h$ as in (3.22), (3.24) in
\cite{LY}. Note that
$\supp\chi\subset\overline{\mathcal B_{1-\sigma}}$. Let
\begin{equation*}
     \chi_k(x) := \chi(|x-R_k|/D_k),
     \qquad
     h_k(x) := h(|x-R_k|/D_k).
\end{equation*}
After scaling and translation,
Proposition \ref{kinloc} yields that for any $0\leq \gamma\leq q$
\begin{equation}\label{eq:stability1}
     \begin{split}
       \tr\gamma(|p+A|-\alpha W) \geq
       & \tr\chi_1\gamma\chi_1(|p+A|-U_{1,\epsilon}^*-\alpha W) \\
       & + \tr(1-\chi_1^2)^{1/2}\gamma(1-\chi_1^2)^{1/2}
       (|p+A|-U_{1,\epsilon}^*-\alpha W) \\
       & - \epsilon^{-1} q \Omega/D_1\,.
     \end{split}
\end{equation}
Here, $U_{1,\epsilon}^*(x) := D_1^{-1} U_\epsilon^*((x-R_1)/D_1)$ and
$\Omega$, $U_\epsilon^*$ were defined in \eqref{eq:omega},
\eqref{eq:uepsilon}. (Note that our $\Omega$ is denoted by $\Omega_1$
in \cite{LY}). Recall that $U_{\epsilon}^*$ and $\Omega$ are independent
of $A$.

We turn to the first term on the right side  of \eqref{eq:stability1}.
Let $C$ be a constant such that
\begin{equation}\label{ass:c}
C \geq (1-\sigma) \left( \alpha \tilde F(|x|) + U_\epsilon^*(x)\right)
\qquad {\rm for\ } |x|\leq 1-\sigma \,.
\end{equation}
Note that $\chi_1$ is supported on a ball of radius $(1-\sigma)
D_1$ centered at $R_1$. Hence $\alpha W(x) = (2/\pi) |x-R_1|^{-1} +
D_1^{-1} \tilde F(|x-R_1|/D_1)$ on the support of $\chi_1$ and we can
apply Theorem~\ref{thm:lth} to obtain the lower bound
\begin{align}\label{lb}\nonumber
\Tr\chi_1\gamma\chi_1(|D-A|-U_{1,\epsilon}^*-\alpha W)
& \geq
\Tr\chi_1\gamma\chi_1\left(|p+A|-\frac{2}{\pi|x-R_1|} -
\frac{C}{(1-\sigma)D_1} \right) \\
& \geq   - 8.4411 \frac{q C^4 }{(1-\sigma)D_1}\,.
\end{align}
We used also that $|\chi_1|\leq 1$.
Inserting \eqref{lb} into \eqref{eq:stability1} we find
\begin{equation*}
     \begin{split}
       \tr\gamma(|p+A|-\alpha W) \geq
       & -q D_1^{-1} \tilde A \\
       & + \tr(1-\chi_1^2)^{1/2}\gamma(1-\chi_1^2)^{1/2}
       (|p+A|-U_{1,\epsilon}^*-\alpha W)
     \end{split}
\end{equation*}
with
\begin{equation*}
     \tilde A := \frac\Omega\eps + 8.4411 \frac{C^4}{(1-\sigma)}\,.
\end{equation*}
This estimate is exactly of the form (3.26) in \cite{LY}, except for
the value of the constant in $\tilde A$ (which is called $A$ in
\cite{LY}). Starting from there one can continue along the lines of
their proof. We need only note that in order to bound the last term in
(3.29) in \cite{LY} one  uses the Daubechies inequality \cite{D},
which holds with the same constant in the presence of a
magnetic field. (This is explained, for instance, in
\cite[Sect.~5]{LLSi}.) We conclude that stability holds as long as
\begin{equation}\label{eq:aj}
\alpha q(\tilde A + J) \leq \frac1{2\pi^2}\,,
\end{equation}
where, as in \cite[Eq.~(3.31)]{LY},
$$
J := 0.0258 \int_{|x|\geq 1-3\sigma} \left[ \frac{2}{\pi|x|} + \alpha
   \tilde F(|x|) + U_\epsilon^*(x)\right]^4\, dx\,.
$$
This completes our proof of Theorem \ref{stability}, except for our
bound on the critical $\alpha$, which we justify now.

As in \cite{LY}, we choose $\sigma=0.3$, $\epsilon=0.2077$ and
$\lambda=0.97$.
Our goal is to prove stability when $q\alpha \leq 1/66.5$. We may assume
$\alpha<1/47$, which is the assumption used in \cite{LY}.
Hence we can use the estimate $J\leq 1.64$ from \cite[Eq.~(3.40)]{LY}.

To bound $\tilde A$, note that $\eps^{-1}\Omega= 0.5571$ as in
\cite[Eq.~(3.30)]{LY}. It remains to choose an appropriate $C$
satisfying (\ref{ass:c}). For $|x|\leq 0.7$ we have $|\tilde
F(|x|)|\leq 1/1.02$. Moreover, for $U_\eps^*$ we use the same estimate
as in \cite{LY}, namely $U_\eps^*(x)\leq 0.2077+ 0.5751 =
0.7828$. Using $\alpha \leq 1/(66.5\, q)\leq 1/66.5$, (\ref{ass:c})
therefore holds with
\begin{equation*}
0.7 \left( 1/(66.5 \cdot 1.02) + 0.7828 \right) < 0.5583 =: C \,.
\end{equation*}
This leads to a value of $\tilde A=1.7287$.
Hence (\ref{eq:aj}) holds for $q\alpha \leq 1/66.5$.


\bibliographystyle{amsalpha}

\end{document}